\documentclass[proof]{WileyASNA-v1}

\articletype{Article Type}%

\received{26 April 2016}
\revised{6 June 2016}
\accepted{6 June 2016}

\begin{document}

\title{Peaked sources and narrow-line Seyfert 1s: a love story}

\author[1,2,3]{Marco Berton}

\author[4]{Emilia Järvelä}

\authormark{Berton \& Järvelä}

\address[1]{European Southern Observatory (ESO), Alonso de C\'ordova 3107, Casilla 19, Santiago 19001, Chile}
\address[2]{Finnish Centre for Astronomy with ESO (FINCA), University of Turku, Vesilinnantie 5, FI-20014 University of Turku, Finland}
\address[3]{Aalto University Mets{\"a}hovi Radio Observatory, Mets{\"a}hovintie 114, FI-02540 Kylm{\"a}l{\"a}, Finland}
\address[4]{European Space Agency (ESA), European Space Astronomy Centre (ESAC), Camino Bajo del Castillo s/n, 28692 Villanueva de la Ca\~nada, Madrid, Spain}

\corres{*M. Berton \& E. Järvelä \\ \email{marco.berton@eso.org};\\ \email{ejarvela@sciops.esa.int}}

%\presentaddress{This is sample for present address text this is sample for present address text}

\abstract{The first similarities between peaked sources (PS) and narrow-line Seyfert 1 (NLS1) galaxies were noticed already twenty years ago. Nowadays, it is known that several sources can share both classifications, and that part of the parent population of $\gamma$-ray emitting NLS1s could be hiding among PS. In this brief review, we describe how and why this orientation-based unification was developed. We also show how the recent discovery of absorbed radio jets in NLS1s, basically invisible at frequencies below 10~GHz, could impact our knowledge of PS and, in particular, render the widely used radio-loudness parameter obsolete. }


%\jnlcitation{\cname{%
%\author{Williams K.}, 
%\author{B. Hoskins}, 
%\author{R. Lee}, 
%\author{G. Masato}, and 
%\author{T. Woollings}} (\cyear{2016}), 
%\ctitle{A regime analysis of Atlantic winter jet variability applied to evaluate HadGEM3-GC2}, \cjournal{Q.J.R. Meteorol. Soc.}, \cvol{2017;00:1--6}.}

%%\fundingInfo{Funding info text.}

\maketitle

%\footnotetext{\textbf{Abbreviations:} ANA, anti-nuclear antibodies; APC, antigen-presenting cells; IRF, interferon regulatory factor}

\newcommand{\kms}{km s$^{-1}$}
\newcommand{\ergs}{erg s$^{-1}$}
\section{Introduction}
\label{sec:intro}
Peaked sources (PS) are a class of powerful ($log P_{1.4\; \rm{GHz}} > 25$ W Hz$^{-1}$) radio sources characterized by a small linear size ($<$15 kpc) of their relativistic jets and by a peaked radio spectrum. The PS population is a mixed bag \citep{Odea21}. Some may be variable radio sources with a temporarily inverted spectrum. Others may be transient sources in which the spectral shape remains the same over time. In this case, intermittent activity of the nucleus may turn the jet on and maintain it for 10$^3$-10$^4$ yrs, and then turn it off and remain quiescent for a longer interval (10$^4$-10$^6$ yrs, \citealp{Czerny09}). Another option is the young age scenario, in which PS represent a population of young sources that are still developing and will eventually grow into classical radio galaxies \citep{Fanti95}. Both of these scenarios have in common the young kinematical age of the relativistic jets. \par
However, PS are not the only class of sources which may have kinematically young jets. Classified not based on their radio spectra as PS, but instead on their optical spectra, narrow-line Seyfert 1 (NLS1) galaxies have by definition a full-width at half maximum of the H$\beta$ emission line lower than 2000 \kms\ \citep{Osterbrock85}. Despite such narrow permitted lines, they are not obscured as Type 2 active galactic nuclei (AGN), but they are instead Type 1 AGN, in which the broad-line region (BLR) and the central engine are directly visible. Although alternative hypotheses exist \citep{Decarli08}, the narrowness of permitted lines is typically attributed to low rotational velocity around a black hole with mass (10$^6$-10$^8$ M$_\odot$, \citealp{Peterson99}) lower than that of other AGN. Since the bolometric luminosity of NLS1s is comparable to that of more massive AGN, they must be accreting close to the Eddington limit \citep[e.g., see][for reviews on their properties]{Komossa18, Gallo18, Lister18, Foschini20}.\par
These properties led to an interesting hypothesis. Let us consider two AGN, located at the same redshift, with the same physical structure and the same life cycle. If one of them has a lower black hole mass it is reasonable to assume that its black hole has spent less time accreting efficiently in an active phase. In this sense, the low-mass AGN can be considered younger than the high-mass one. This could be the case for NLS1s, as they may be the progenitors of AGN with higher black hole masses \citep{Mathur00}, and more recent statistical analysis seem to confirm this scenario \citep{Fraixburnet17a, Jarvela17}. \par
Some NLS1s harbor powerful relativistic jets that, when closely aligned with the line of sight (i.e., beamed), can be sources of $\gamma$-ray emission \citep{Abdo09c}. Such jetted NLS1s aligned with our line of sight seem to share most of the properties of their non-jetted counterparts, such as black hole mass and Eddington ratio \citep{Berton15a}, and for this reason they could be the progenitors of other $\gamma$-ray sources like flat-spectrum radio quasars (FSRQs, \citealp{Foschini15}). However, it is known that, when observed at large angles, FSRQs appear as high-excitation radio galaxies (HERGs, \citealp{Urry95, Giommi12}). We mentioned before that some PS sources could be young radio galaxies. Therefore, some of these genuinely young PS may be the misaligned counterpart (i.e., the parent population) of $\gamma$-ray NLS1s \citep{Berton17}.

\section{A new unified model?}
The similarity between some PS and jetted NLS1s was already noticed twenty years ago, when the powerful radio source PKS 2004-447 was identified as a PS with an NLS1 optical spectrum \citep{Oshlack01}. After that, a number of sources sharing both classifications have been found \citep{Komossa06, Yuan08, Gu15, Liao20, Yao21}. Based on this, \citet{Berton16c} proposed that a subclass of PS called low-luminosity compact (LLC) sources \citep{Kunert10a} could be good candidates as part of the parent population of $\gamma$-ray NLS1s. Some of these LLCs can be optically classified as Type 2 HERGs \citep{Kunert10b}, and their black hole mass and Eddington ratio distributions overlap with those of beamed NLS1s. \citet{Berton16c} also studied their radio luminosity function (LF). This technique allows to add a model of relativistic beaming to the LF of an unbeamed population, which can be then compared to the observed LF of the beamed population \citep{Padovani92}. This analysis showed that LLC can belong to beamed NLS1s' parent population if the diffuse radio emission of the latter is negligible with respect to their non-thermal core emission. As later confirmed by \citet{Berton18a}, this is exactly what is observed among beamed NLS1s that, unlike FSRQs \citep{Antonucci88}, typically lack diffuse emission. \par
A potential issue with this unification is that, while jetted NLS1s tend to be harbored in spiral galaxies
\citep{Berton19a, Jarvela18, Kotilainen16, Olguiniglesias20, Hamilton21}, PS are usually hosted by ellipticals \citep{Odea21}. However, in general, HERG-like PS tend to have bluer colors than LERG, possibly pointing toward a more gas-rich host \citep{Heckman14}. Furthermore, to the best of our knowledge no dedicated study has been carried out for LLC/HERGs. Since their central engine properties are similar to those of jetted NLS1s, it would not come as a surprise to find that they also are hosted by spirals. This point, anyway, clearly shows that jetted NLS1s cannot be unified with the whole population of PS. Only a very specific subset of them can be part of the parent population of jetted NLS1s, very likely those with a relatively low-mass black hole and a high Eddington ratio. \par
As previously mentioned, there is no doubt that NLS1s and some PS do share some properties. The two classifications are not mutually exclusive, since they are based on different frequencies. Several $\gamma$-NLS1s show peaks above 10~GHz, similarly to what is observed in some PS \citep{Lahteenmaki17}. Furthermore, radio observations carried out with very long baseline interferometry \citep[e.g.,][]{Schulz15, Caccianiga17} showed that some jetted NLS1s respect the criteria to be classified, in radio, as PS. Two of these jetted NLS1s with PS classification, PKS 2004-447 and 3C 286 \citep[e.g.,][]{Yao21, Berton21b}, are also $\gamma$-ray sources included in the \textit{Fermi} 4FGL. Since this catalog only includes six PS \citep{Abdollahi20}, and two of them are NLS1s, it is clear that sources with both classifications are potentially very interesting candidates as high-energy emitters. 

\section{The riddle of invisible jets}

The most interesting discovery of the last decade in the NLS1 field \citep{Foschini20} is the detection at 37~GHz of a few radio-quiet or even radio-silent (i.e. never detected in any radio survey) NLS1s with the Mets\"ahovi Radio Telescope \citep{Lahteenmaki18}. Seven of these objects were unexpectedly found flaring multiple times at Jy-level with relatively short timescales. The only reasonable mechanism to produce this kind of emission is by means of a relativistic jet, as seen in blazars. Follow-up observations with the VLA revealed something even more unexpected. These NLS1s all have very faint emission at lower radio frequencies. Between 1.4 and 10~GHz, their flux densities are of the order of $\sim$mJy or, more often, $\mu$Jy \citep{Berton20b}. The beam sizes of VLA and Mets\"ahovi are very different. This, however, cannot account for such a variable emission, as it would be located far from the nucleus or even outside of the host galaxy. Furthermore, new observations are revealing that radio flares correspond to a brightening of the nucleus at other frequencies (Romano et al., in prep). This indisputably proves that the flares are produced by the NLS1 nuclei. What can cause such a peculiar behavior? \par
The spectrum of these objects is dominated by a power law with spectral index $\alpha_\nu \sim -0.7$ ($S_\nu \propto \nu^\alpha$) below 10~GHz. Above this frequency, instead, the spectrum is rising with a spectral index up to $\alpha_\nu = 6.8$. The emission at high frequency was measured only during a flaring state of the jet, therefore this number is definitely an upper limit. However, it is evident that the spectrum cannot have the same spectral index measured at low frequencies up to $\sim$37~GHz, or these flares would lead to an unrealistic increase in the flux density of $>$4500 times. \par
An inverted radio spectrum is nothing new among PS, as it is known that small jets are peaking at relatively high frequencies. But in our case, the peak would be around 40~GHz, which would be unprecedented, and also short lived, because it would move to much lower frequencies in a few years. A more likely explanation for the inverted spectrum we observe is that the origin of the radio emission in the two frequency domains is different. At low frequency the jet is not visible, and the radio emission is likely produced by a combination of nuclear activity and star formation (e.g., supernova remnants), which is often very strong in NLS1s \citep{Sani10}. The high frequency emission, instead, originates in a small-scale relativistic jet, possibly closely aligned with the line of sight as in blazars. We hypothesize that a dense region of ionized gas is present between the star forming regions and the small-scale jet. The ionization source may be either young hot stars or a bow shock produced by the jet itself. The gas is optically thick at low frequencies because of free-free absorption. For this reason, only the emission from the star formation, which is occurring farther away, is visible at low frequencies. At high frequency, instead, the ionized gas is optically thin, and the jet radio emission can escape freely. The relativistic jet must have a relatively small scale, otherwise the size and density of the region of ionized gas would be unrealistically large. This may be in agreement with the young age scenario for NLS1s. A small-scale jet, indeed, may have formed only rather recently. \par

\section{The death of radio-loudness}

The discovery of absorbed relativistic jets needs confirmation by means of multiwavelength observations. If our hypothesis is correct, however, it could have severe implications, especially regarding the use of the radio loudness parameter. Defined by \citet{Kellermann89}, it is the ratio between optical and radio flux densities, $R = S_{\rm opt}/S_{\rm radio}$. However, as suggested already in \citet{Padovani17} and \citet{Jarvela17}, AGN should be classified based on their real physical properties, not based on an artificial parameter with an arbitrary threshold. The radio loudness parameter had its use when our knowledge of AGN and different AGN-related phenomena was still limited, and most studies focused on either very bright and distant quasars, or nearby, only moderately luminous Seyferts. Due to this bias in interests the so-called radio loudness dichotomy was formulated \citep{Cirasuolo03}, stating that there are two intrinsically different populations of AGN: radio-loud and radio-quiet sources, and no smooth transition between the two populations. In this scheme the high radio-loudness was associated with bright AGN with relativistic jets, and the radio-quiet sources were thought to be mostly non jetted. During the past decades it has become increasingly clear that the radio-loudness distribution of AGN is not bimodal \citep[e.g.,][]{Jarvela15}, and this view of AGN is too simplistic. 

Already 20 years ago \citet{Ho01} showed that when carefully measuring the radio-loudness of the nuclear component in radio-quiet Seyfert galaxies, majority of them would actually be classified as radio-loud. This highlights one of the issues with the radio-loudness parameter: the host galaxy. In nearby sources the host galaxy can considerably contribute to the optical light measured from the galaxy, and decrease the radio-loudness of the source. This aspect is not taken into account in the original definition of the radio-loudness parameter. The host galaxy can also have an opposite effect, as shown in \citet{Caccianiga15}. They study a sample of radio-loud NLS1s and conclude based on their mid-infrared properties that actually some of these sources might appear radio-loud due to the enhanced star formation processes in the host galaxy.

The radio-loudness parameter is too ambiguous to be reliable, or even useful. In addition to the issues mentioned above, its value also depends on, for example, the optical and radio variability of AGN, the instruments used on either band, and their sensitivity. The final nail to the coffin of radio-loudness was the discovery of most likely relativistic jets in sources that were previously classified as radio-quiet or radio-\textit{silent}. For now these sources are a curiosity, but we do not know how common these or similar properties among AGN are. However, this example perfectly highlights the poor predicting power of the radio-loudness parameter, and it can be easily seen how using it might lead to misleading or even wrong results, especially among sources, such as NLS1s, where different components can co-exist. Using such a parameter to, for example, divide samples in statistical studies would obviously skew the results, and lessen their reliability and significance. These problems can be avoided by properly classifying the sources based on their physical properties, instead of a vague parameter, repeatedly proven to be misleading. The radio-loudness parameter truly seems to be a relic of the past, and we thus kindly suggest that we finally lay it to rest.

\bibliography{biblio}%
\section*{Authors' Biography}
\begin{biography}{}
{\textbf{Marco Berton} is an ESO fellow, previously working at the Finnish Centre for Astronomy with ESO. He obtained his PhD from University of Padova in 2016. His research interest are active galaxies, and particularly NLS1s. \textbf{Emilia J\"arvel\"a} is an ESA research fellow, previously employed at University of California, Santa Barbara. She obtained her PhD from Aalto University Mets\"ahovi Radio Observatory in 2018. Her research interests encompass the evolution of active galaxies, NLS1s, and other young AGN.}
\end{biography}

\end{document}